\documentclass[10pt,conference]{IEEEtran}
\usepackage{amsmath,amsfonts,bm,graphicx}
\begin{document}

\title{Instanton  analysis
of Low-Density-Parity-Check codes in the error-floor regime}

\author{\authorblockN{Mikhail Stepanov}
\authorblockA{T-13 and CNLS, Theoretical Division,\\
Los Alamos National Laboratory,\\
Los Alamos, NM 87545, USA\\
\& Institute of Automation and Electrometry,\\
Novosibirsk 630090, Russia\\
Email: stepanov@cnls.lanl.gov}
\and
\authorblockN{Michael Chertkov}
\authorblockA{T-13 and CNLS, Theoretical Division,\\
Los Alamos National Laboratory,\\
Los Alamos, NM 87545, USA\\
Email: chertkov@lanl.gov}}

\maketitle

\begin{abstract}
  In this paper we develop instanton method introduced in
  \cite{04CCSV,05SCCV,05SC} to analyze quantitatively performance of
  Low-Density-Parity-Check (LDPC) codes decoded iteratively in the
  so-called error-floor regime. We discuss statistical properties
  of the numerical instanton-amoeba scheme
  focusing on detailed analysis and comparison of two
  regular LDPC codes: Tanner's $(155, 64, 20)$ and
  Margulis' $(672, 336, 16)$ codes. In the regime of moderate values
  of the signal-to-noise ratio we critically compare results of the
  instanton-amoeba evaluations against the standard
  Monte-Carlo calculations of the Frame-Error-Rate.
\end{abstract}

\section{Introduction}

Low-Density-Parity-Check (LDPC) codes \cite{63Gal,99Mac} are
special, not only because they can approach virtually error-free
transmission limit, but mainly because a computationally efficient
iterative decoding scheme is readily available. The approximate
scheme is however incapable of matching performance of the
Maximum-Likelihood (ML) decoding beyond the so-called error-floor
threshold \cite{04Ric} found at higher Signal-to-Noise-Ratios (SNR),
$s$. Moreover, to estimate the error-floor asymptotic in modern
high-quality systems is an important but notoriously difficult task
\cite{04Ric,05VK}. The major problem here is due to the standard
direct numerical methods, of the Monte-Carlo (MC) type, inability to
determine BER below $10^{-9}$. Recently, we (in collaboration with
V.~Chernyak and B.~Vasic) have proposed a physics inspired approach
that is capable of a computationally tractable analysis of the error
floor phenomenon \cite{04CCSV,05SCCV,05SC}. We proposed an efficient
numerical scheme (coined instanton-amoeba), which is ab-initio by
construction, i.e.\ the optimization scheme required no additional
assumptions. The scheme is also generic, in that there are no
restrictions related to the type of decoding or channel. The
instanton-amoeba scheme was first tested for loop-less models
\cite{04CCSV}, then applied to realistic codes with loops in the
Additive-White-Gaussian-Noise (AWGN) channel \cite{05SCCV}, and also
to analysis of the Laplacian channel \cite{05SC}.

In this paper we continue the approach and analyze two popular codes
by means of the instanton-amoeba technique. Our aim is first of all
to quantify utility of the instanton-amoeba scheme for different
codes. We characterize performance of the scheme discussing
statistical distribution of the instanton's length found in the
result of multiple attempts of the instanton-amoeba. We also focus
on testing and explaining performance of the method at moderate
values of SNR. We discuss convergence of the Monte-Carlo results to
the instanton asymptotics observed for different codes and different
number of iterations.

We find that the large SNR behaviors shown by the two codes are
qualitatively different. One important difference is coming from the
fact that the instanton with the lowest effective distance (that is
the single most damaging configuration of the noise) corresponds to
a pseudo-codeword in the case of the $(155, 64, 20)$ code,  while
the instanton is a valid codeword in the case of the $(672, 336,
16)$ code. We observe that for both codes there are many other (than
minimal) instanton solutions with distance higher than of the
minimal instanton. However,  characterizing the spectra of these
``excited" instantons with the distance higher than minimal we find
that while in the case of the $(155, 64, 20)$ code there are many
``excited" instantons with distances very close to the minimal one,
the ``excited" instantons of the $(672, 336, 16)$ code are separated
by a large gap from the minimal one. Moreover, the phase volume in
the phase (noise) space associated with the minimal instanton
(correspondent to a codeword) in the case of the $(672, 336, 16)$
code is relatively small,  so that one needs to make special
(biasing) efforts to force the instanton-amoeba to find the minimal
instanton. This suggests that the actual performance of the $(672,
336, 16)$ code at moderate SNR is essentially better than one would
anticipate based on the Hamming distance of the code (equal to $16$)
as the asymptote is mainly defined by the pseudo codewords above the
gap. Contribution of the ``excited" instantons dominates FER at some
moderate values of SNR while the codeword contribution eventually
takes over at higher values of SNR. These conclusions follow from
the instanton-amoeba analysis complemented and contrasted against
the standard MC analysis.

We also perform analysis of the FER dependence on the number of
decoding iterations for the two codes.  The analysis reveals the
same basic phenomenon: the largest SNR asymptotic is controlled by
the minimal length instanton, however a nontrivial interplay between
different instantons may affect an intermediate SNR asymptotic. The
major effect at the transient, intermediate SNR, state  can be
related to competition between the contribution to FER from the
minimal distance instanton and some higher distance instantons
occupying a larger volume of the noise space.

\section{Notations}

\begin{figure}[tb]
  \centerline{\includegraphics[width=3in]{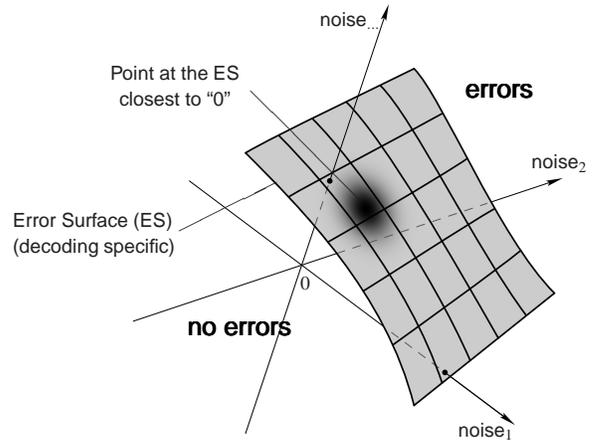}}
  \caption{Illustration for the instanton method. The noise space is
    divided into areas of successful and erroneous decoding by error
    surface. The point at the error surface closest (in the appropriate
    metrics) to the point of zero noise is
    the most probable configuration of the noise causing the decoding error.
    Contribution from the special configuration of the noise, the instanton,
    and its close vicinity estimates the noise integral for FER.}
  \label{inst}
\end{figure}

Sending a codeword ${\bm\sigma}=\{\sigma_i=\pm 1; i=1,\cdots,N\}$
into a noisy channel results with the probability $P({\bm x}|{\bm
\sigma})$ in corruption of the original signal, ${\bm
x}\neq{\bm\sigma}$. The decoding goal is to infer the original
message from the received output, ${\bm x}$. Assuming that coding
and decoding are fixed one studies Frame-Error-Rate (FER) to
characterize performance of the scheme,
%\begin{eqnarray}
 $\mbox{ FER} = \int d{\bm x} \,\, \chi_{\mbox{\scriptsize error}}({\bm
   x}) P({\bm x}|{\bm 1})$,
  %\label{integral}
%\end{eqnarray}
where $\chi_{\mbox{\scriptsize error}} = 1$ if an error is detected
and $\chi_{\mbox{\scriptsize error}} = 0$ otherwise. In symmetric
channel FER is invariant with respect to the original codeword, thus
all-$(+1)$ codeword can be assumed for the input. When SNR is large
FER as an integral over output configurations is approximated by,
\begin{eqnarray}
\mbox{FER}\sim \sum\limits_{\mbox{\scriptsize inst}}
V_{\mbox{\scriptsize inst}} \times P\left({\bm x}_{\mbox{\scriptsize
inst}}|{\bm 1}\right), \label{FER}
\end{eqnarray}
where ${\bm x}_{\mbox{\scriptsize inst}}$ are the special instanton
configurations of the output maximizing $P({\bm x}|{\bm 1})$ under
the $\chi_{\mbox{\scriptsize error}} = 1$ condition, and
$V_{\mbox{\scriptsize inst}}$ combines combinatorial and
phase-volume factors. See Fig.~\ref{inst} for illustration.
Generically, there are many instanton configurations that are all
local maxima of $P({\bm x}|{\bm 1})$ in the noise space. Individual
contributions into FER decrease significantly with SNR increase. At
large SNR only instanton with the highest $P({\bm
x}_{\mbox{\scriptsize inst}}|{\bm 1})$ is relevant.

AWGN channel is defined by,
\begin{eqnarray} \label{AWGN}
 P({\bm x}|{\bm \sigma}') \!=\! \prod\limits_i p(x_i|\sigma_i'), \ \
 p(x|\sigma) \!\propto\! \exp\left(\!-\frac{(x-\sigma)^2s^2}{2}\! \right).
\end{eqnarray}
If the detected signal at a bit is $x$, the respective
log-likelihood at the bit is $h = \ln(p(x|1)/p(x|-1)) / 2s^2 = x$,
where min-sum decoding is assumed and one chooses to measure it in
the units of SNR squared, $s^2$. For the AWGN channel finding the
instanton means minimizing $l^2 = \sum_i (1 - x_i)^2$, with respect
to the noise vector ${\bm 1} - {\bm x}$ in the $N$-dimensional
space, under the condition that the decoding terminates with an
error. Instanton estimation for FER at the highest SNR, $s \gg 1$,
is $\sim \exp(-l_{\mbox{\scriptsize inst}}^2 \cdot s^2/2)$, while at
moderate values of SNR many terms from the right-hand-side of
Eq.~(\ref{FER}) can contribute to FER comparably.

\section{Numerical scheme. Details and Validation.}

In our numerical scheme the length $l_{\mbox{\scriptsize inst}}$ was
found by a downhill simplex method \cite{65NM}, also called
``amoeba'', with accurately tailored (for better convergence)
annealing. We repeat the instanton-amoeba evaluation many times,
always starting from a new set for initial simplex chosen randomly.
$l$,  as a function of noise configuration inside the area of
unsuccessful decoding, has multiple minima each corresponding to an
instanton. Multiple attempts of the instanton-amoeba evaluations
gives us not only the instanton with the minimal
$l_{\mbox{\scriptsize inst}}$ but also the whole spectra of higher
valued $l_{\mbox{\scriptsize inst}}$.

We develop two different versions of ``amoeba",  ``soft" and
``hard". In ``soft amoeba" the minimization function decreases with
noise probability density in erroneous area of the noise, while in
area of successful decoding the function is made artificially big
(to guarantee that the actual minimum is achieved inside the
erroneous domain). In the ``hard amoeba'' case minimization is
performed only over all orientations of the noise vector, while the
length of the vector corresponds exactly to respective point at the
error surface, that is the surface separating domains of errors from
the domain of correct decoding. (See Fig.~\ref{inst} for
illustration.) This special point at the error surface is found
numerically by bisection method.

In \cite{05SCCV,05SC} the ``hard amoeba" was used. Preparing
material for this paper we found that even though the ``hard amoeba"
outperforms the ``soft amoeba" for relatively short codes,  the
later one has clear advantage in the computational efficiency for
mid-size and long codes.   Thus,  the results presented in the paper
were derived primarily by means of the ``soft amoeba".

Once found numerically the validity of the instanton solution could
be checked against a theoretical evaluation. This theoretical
approach, introduced in \cite{05SCCV,05SC}, is based on the notion
of the computational tree (CT) of Wiberg \cite{96W} built by
unwrapping the Tanner graph of a given code into a tree from a bit
for which one determines the probability of error. The concept of CT
is useful because the result of iterative decodings at a bit of an
LDPC code and at the tree center of the respective CT are equal by
construction \cite{96W}. The initial messages at any bit of the tree
are log-likelihoods and, therefore, the result obtained in the tree
center is a linear combination of the log-likelihoods with integer
coefficients, so the error surface condition becomes $\sum_i n_i h_i
= 0$ with integer $n_i$ that depend on CT structure. For AWGN
channel the instanton length is equal to $l_{\mbox{\scriptsize
inst}}^2 = (\sum_i n_i)^2/(\sum_i n_i^2)$ \cite{96W}. The definition
of $n_i$ was generalized in \cite{05SCCV}. %, where it was shown that
%individual bits on CT could contribute $-1$ or even not an integer
%number to corresponding $n_i$.
In spite of its clear utility
%for analysis of decoding with relatively small number of iterations
the CT approach becomes impractical for larger number of iterations.
Thus, in this paper discussing (among other issues) dependence of
FER on the number of iterations, we use the CT approach only to
verify validity of the instanton-amoeba results for relatively small
number of iterations.

Let us also mention difficulties  we encountered extending the
instanton-amoeba scheme for large number of iterations. First of
all, increase of $N_{\mbox{\scriptsize it}}$ simply means longer
computations. The other more important effect is associated with
enhancement of irregular, stochastic component in decoding observed
with $N_{\mbox{\scriptsize it}}$ increase. One finds that already a
slight variation in the noise can drastically change results.
Sensitivity of the instanton-amoeba scheme and dynamical decoding to
small variations of the noise will be discussed in details elsewhere
\cite{SC}.

\begin{figure}[tb]
  \centerline{\includegraphics[width=3in]{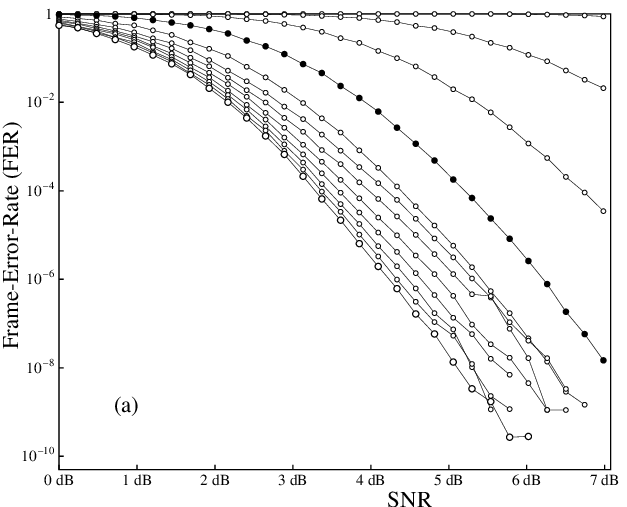}}
  \medskip
  \centerline{\includegraphics[width=3in]{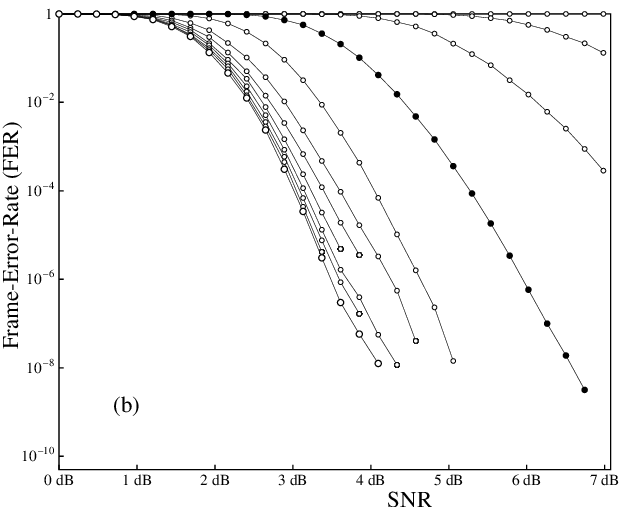}}
  \caption{MC evaluation of FER vs SNR (one uses standard in
    error-correction literature notations in dB: $\mbox{SNR(dB)} = 20
    \log_{10} s$) for $(155, 64, 20)$ (top, (a)) and $(672, 336, 16)$
    (bottom, (b)) codes performed over AWGN channel and decoded by min-sum
    algorithm. Different curves correspond to decoding with $0$ (no
    decoding), $1$, $2$, $4$, $8$, $16$, $32$, $64$, $128$, $256$, $512$
    and $1024$ iterations. Filled dots for the data points correspond to $4$ iterations decoding.
    Improvement of the code performance with the
    number of iterations is monotonic.}
  \label{bere}
\end{figure}

\begin{figure}[tb]
  \centerline{\includegraphics[width=3in]{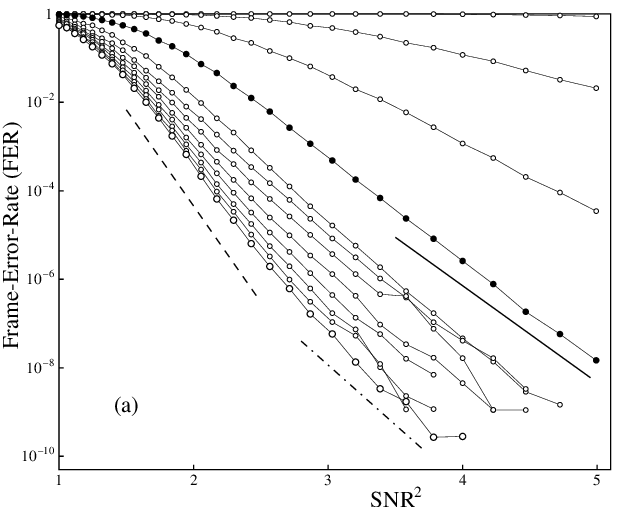}}
  \medskip
  \centerline{\includegraphics[width=3in]{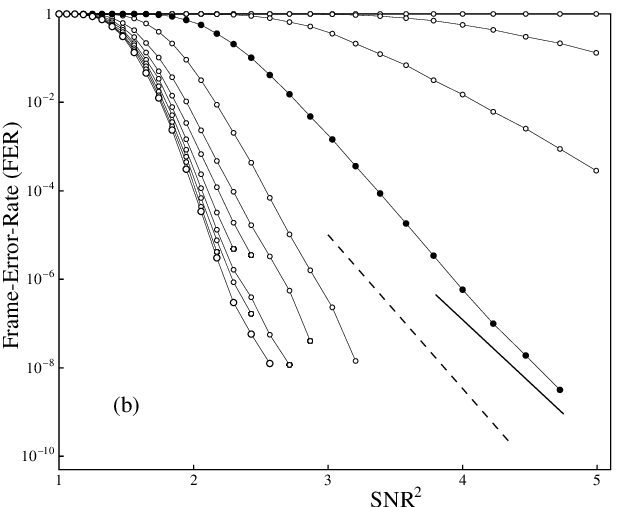}}
  \caption{The same data as shown in Fig.~\ref{bere}
    are replotted here as FER vs
    SNR$^2$. Dashed line corresponds to the Hamming distance
    asymptotics, $\mbox{FER} \sim \exp(-d_{\mbox{\scriptsize min}} \cdot
    s^2/2)$. Solid line corresponds to the minimal distance instanton for $4$ iterations, $\mbox{FER} \sim
    \exp(-l^2_{\mbox{\scriptsize min}} \cdot s^2/2)$, where
    in the case (a) of the $(155,64,20)$ code, $l^2_{\mbox{\scriptsize min}} = 46^2/210$ and
    in the case (b) of the $(672,336,16)$ code
    $l^2_{\mbox{\scriptsize min}} = 46^2/162$. Dot-dashed line in Fig.~(3b) corresponds
    to the special instanton configuration described by
    Fig.~\ref{conf_4_400} which withstands $400$ iterations.}
  \label{ber}
\end{figure}

Finally, we also use standard MC approach for comparison with the
instanton-amoeba analysis in the regime of moderate SNR values,
where on one side FER can still be accessed by MC, but on the other
side the instanton asymptotic already applies. This allows us to
study convergence of the Monte-Carlo data to the instanton
asymptotics,  and thus to explore possible effect of other higher
distance instantons and also effect of their phase space (volume)
factors.

\section{Instanton-amoeba analysis of the Tanner $(155,64,20)$ and
Margulis $(672,336,16)$ codes:  Results and Discussions.}

We consider two regular LDPC codes: $(155,64,20)$ code of Tanner
\cite{01TSF} and $(672, 336, 16)$ code of Margulis \cite{82M,02MP}
with the prime number $p = 7$. The minimal Hamming distance of the
codes is $d_{\mbox{\scriptsize min}} = 20$ and $d_{\mbox{\scriptsize
min}} = 16$ respectively. This translates into the following largest
SNR expectation for FER if the decoding is ML: $\sim \exp(-d_{\rm
min} \cdot s^2/2)$. The decay of FER with SNR at the largest SNR is
not that steep if an approximate iterative scheme is used, $\sim
\exp(-l_{\mbox{\scriptsize inst}}^2 \cdot s^2/2)$ where thus
$l_{\mbox{\scriptsize inst}}^2 \le d_{\rm min}$.

The results of the MC simulations for the two selected codes are
shown in Fig.~\ref{bere}, \ref{ber}, where $\log(\mbox{FER})$ and
$\mbox{SNR}^2$ variables are used for plotting in Fig.~\ref{ber}.
Notice that the variables used in Fig.~\ref{bere} are standard for
error-correction literature, while the variables used in
Fig.~\ref{ber} are more appropriate for purposes of the asymptotic
analysis as the instanton asymptotic, ${\rm FER} \propto
\exp(-l_{\mbox{\scriptsize inst}}^2 \cdot s^2/2)$, becomes a
straight line in these later variables.

Our analysis suggests that the leading $s\to\infty$ asymptotic for
FER is governed by the instanton with the lowest length found, i.e.
${\rm FER} \sim \exp(-l_{\mbox{\scriptsize lowest inst}}^2 \cdot
s^2/2)$. We have checked this prediction against direct MC
simulations and found that the instanton asymptotic either sets
already in the range accessible to the MC simulations, ${\rm
FER}<10^{-9}$, or otherwise one observes a tendency for the MC curve
to change its slope towards the instanton asymptotic. The former
case is clearly seen on example of the $(155,64,20)$ code with $4$
iterations (the respective asymptotic is shown in Fig.~\ref{ber} as
the dashed line). In this (and some other cases) the instanton
approximation starts to work well already at moderate values of SNR.
Explanation for this effect is given in \cite{05SC}.

\begin{figure}[t]
  \centerline{\includegraphics[height=3in,angle=270]{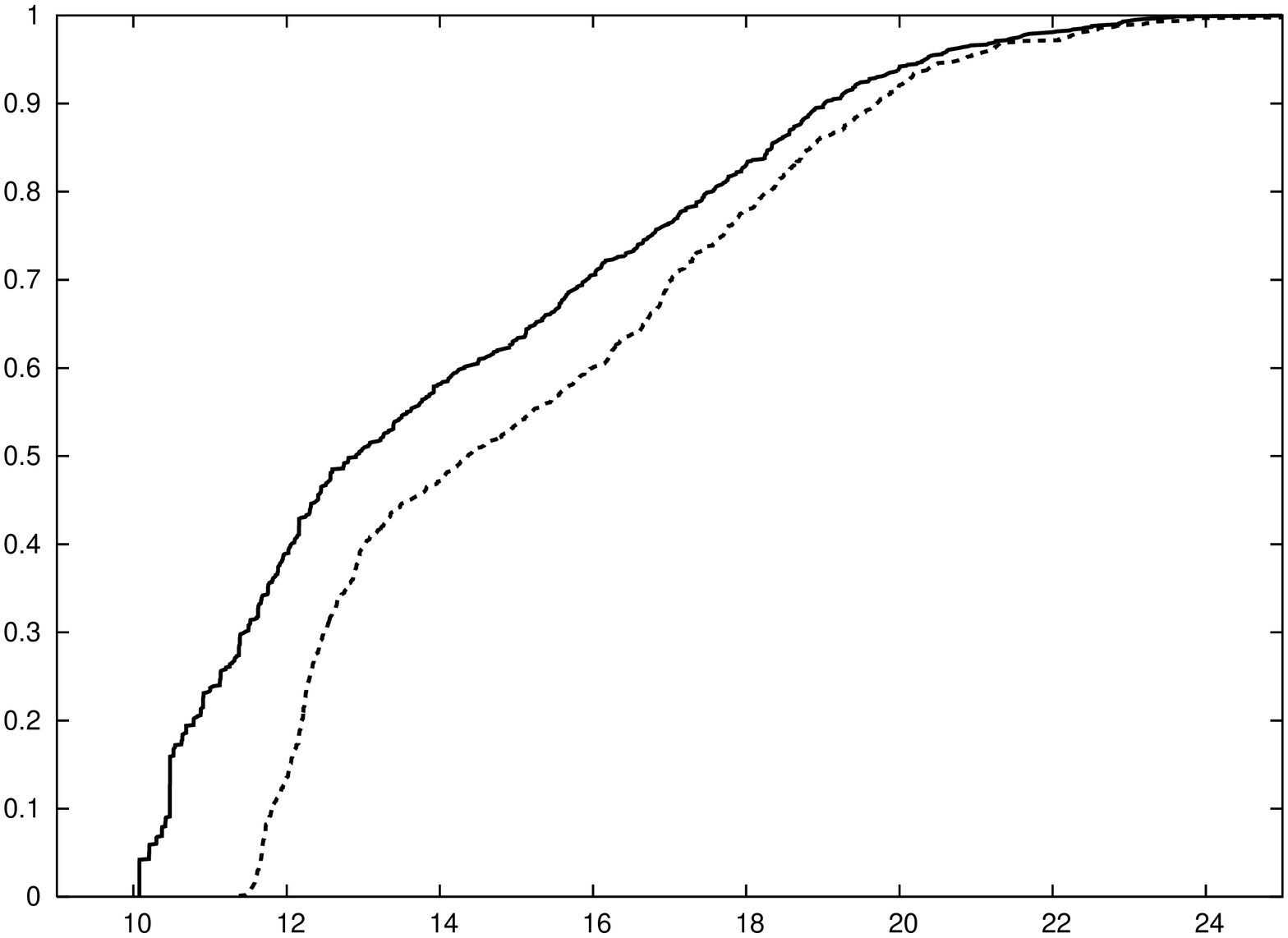}}
  \medskip
  \centerline{\includegraphics[height=3in,angle=270]{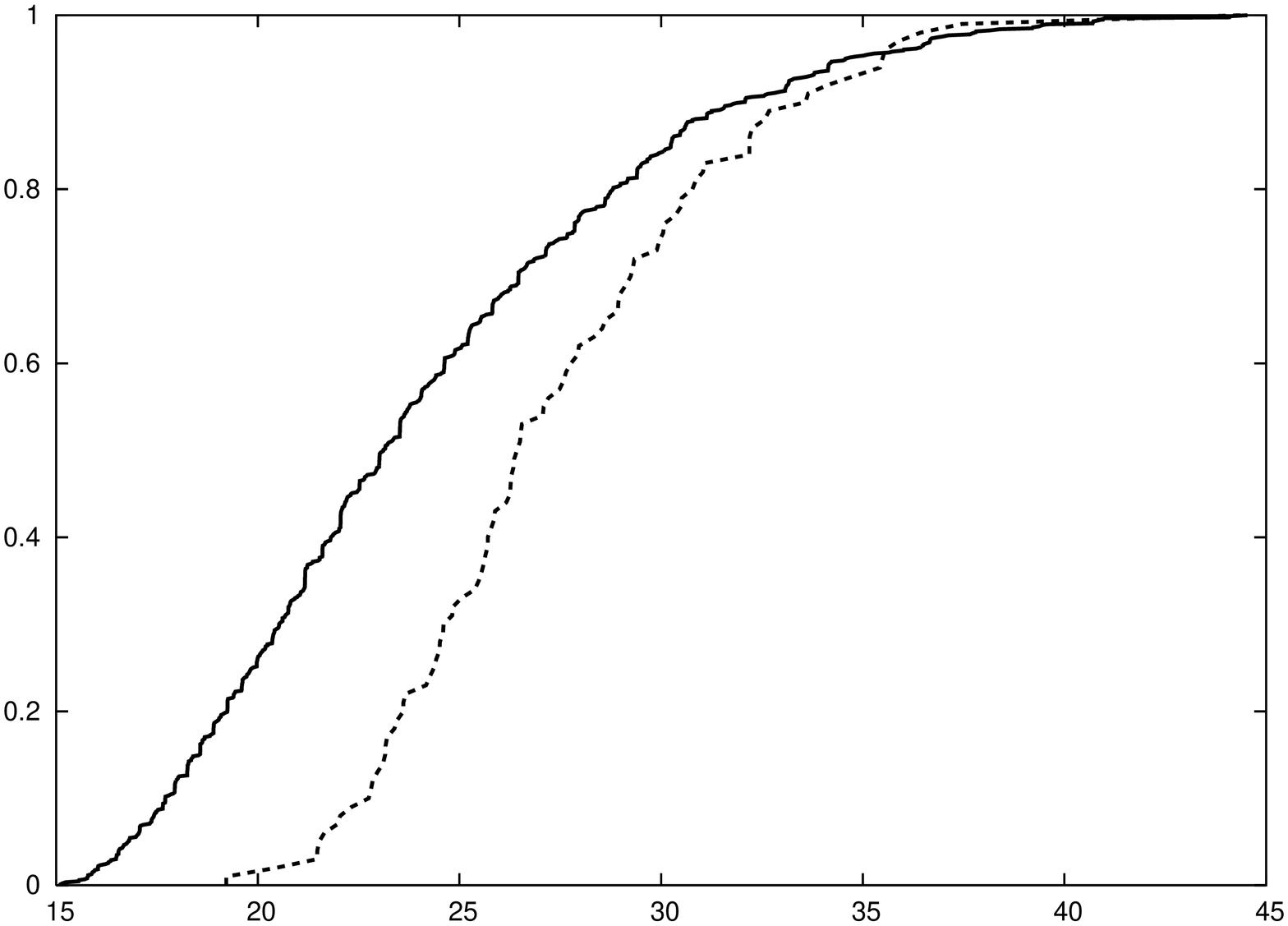}}
  \caption{Distribution function of $l_{\mbox{\scriptsize inst}}^2$ for
    4- and 8-iteration decoder (solid and dashed curves respectively)
    obtained by instanton-amoeba scheme.
    Plot on the top describes 1000 attempts of the instanton-amoeba scheme
    for $(155, 64, 20)$ code.  Plot on the bottom shows data for Margulis' $(672, 336, 16)$ code
    collected in $863$ and $100$ attempts
    (for $4$ and $8$ iterations respectively) of the instanton-amoeba scheme.}
  \label{fl2}
\end{figure}

Describing Fig.~\ref{ber} for both codes we observe that the rate of
FER decrease with SNR at its moderate values (correspondent to ${\rm
FER}\sim 10^{6\div 8}$) increases with the number of iterations.  On
the other side one also finds that the effective slope of the FER
dependence on SNR at this moderate values is essentially larger than
the one correspondent to the minimal instanton which controls the
largest SNR asymptotic. We interpret this as effect of the volume
factor in Eq.~(\ref{FER}). Indeed, higher effective distance
instantons have larger volume factors. (This observation is
indirectly confirmed by Fig.~\ref{fl2} showing that the instantons
with higher effective distance are more probable.) Thus, at moderate
SNR effect of the volume factor becomes essential shifting the
transient asymptotic towards higher effective distance. This
asymptotic could be even steeper than the one corresponding to the
Hamming distance of the code. However, and in spite of the large
volume factor that is not changing much with SNR, these large phase
volume but high effective distance contributions to the right hand
side of Eq.~(\ref{FER}) will become less and less important with SNR
decrease. Such behavior will always be seen if the lowest instantons
or codewords have small phase volume, as observed for Margulis'
$(672, 336, 16)$ code (see Fig.~\ref{ber}(b)). Interesting
consequence of this volume factor effect is that in this regime of
moderate SNR the $(672, 336, 16)$ code performs better than one
would expect knowing a relatively low value of its Hamming distance
$d_{\mbox{\scriptsize min}} = 16$. Looking at Fig.~\ref{fl2}b
describing statistics of multiple random attempts of the
instanton-amoeba scheme for the $(672, 336, 16)$ code one also finds
no instantons with the effective weight lower than $19$ (for $8$
iterations). These low distance instantons can still be found if
initial configuration for instanton-amoeba is carefully
pre-selected. The special initial configuration can be either a
codeword with the Hamming distance $16$ or configuration found by
the LP-loop method described in \cite{CSLP}.

\begin{figure}[t]
  \centerline{\includegraphics[width=3in]{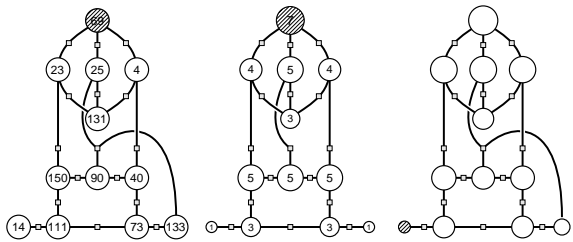}}
  \caption{Figure describes a special instanton configuration for the
  Tanner $(155,64,20)$ code that shows remarkable stability both in
  the structure and in the effective length. Left panel defines the
  structure with bits numbered according to the rule explained in
  \cite{05SCCV}. Middle and right panels show the instanton
  configurations found for $4$ and $400$ iterations with the area
  of the circles (and the integer values of the middle panel) correspondent to the
  noise amplitude. Erroneous bits (for $4$ and $400$ iterations respectively) are shaded.}
  \label{conf_4_400}
\end{figure}

One also notes, continuing to discuss Fig.~\ref{fl2}, that amoeba
frequently finds the optimal instanton for $(155, 64, 20)$ code. The
situation however is drastically different for $(672, 336, 16)$ code
when amoeba explores mainly instantons with relatively high
effective distance $l_{\mbox{\scriptsize inst}}^2 > 16$. Another
interesting observation one can draw from Fig.~\ref{fl2} concerns
the step-like dependence of the probability on $l_{\mbox{\scriptsize
inst}}^2$. These steps correspond to the instantons with large
volume factors.

Finally we show in Fig.~\ref{conf_4_400} an instanton configuration
for the $(155,64,20)$ code that survives $400$ iterations. This
instanton was found through special biasing efforts. One identifies
$12$ (of the total number of $155$) bits, associated with the lowest
distance instanton found by standard procedure for $4$ iterations,
and restricts amoeba to varying noise value at the $12$ bits only
while imposing exact zero for the noise at the other bits. The
result of this restricted minimization was later used as a starting
point for the full amoeba. This way we show that even for this very
large number of iterations the instanton distance does not exceed
the value of $12.45$. Slope correspondent to this number is shown as
a dot-dashed line in Fig.~\ref{ber} for the Tanner code. One finds a
good agreement between this single instanton asymptotic and the MC
results for large number of iterations. This approach, of
restricting the instanton-amoeba to variations only in small subset
of bits, applied to the $(672, 336, 16)$ code also gives an
impressive result. One finds instanton of the distance $\approx
14.48$ (i.e. of the distance smaller than the Hamming distance of
the code) that survives $100$ iterations. For this instanton noise
is non-zero only on $16$ bits (correspondent to the Hamming distance
codeword). Detailed analysis and dynamical description of this
special configurations surviving anomalously large number of
iterations will be given in \cite{SC}.

\section{Other codes}

The instanton-amoeba approach is generic in the sense of its
potential utility for variety of codes, decoding schemes and channel
models. The two codes discussed in this text got most of our
attention, however we also analyzed (and are actually continuing
doing so) other interesting codes. The analysis of these other codes
is not as complete and detailed as of the two selected codes. Thus,
in what follows, we make  few not really coherent remarks about the
current stage of the other codes exploration.

One interesting long code is of Margulis type \cite{82M,02MP} with
$p=11$. This code consists of $2640$ bits and $1320$ checks and is
the largest code we tested with the instanton-amoeba so far. Hamming
distance of the code is not known, while a bound of $220$ is cited
in \cite{02MP}. The lowest length instanton configuration we found
(in not very exhausting search) by the instanton-amoeba for $8$
decoding iterations gives $l^2 \approx 80.879$. We expect existence
of significantly lower length instantons, as the lowest distance
configuration of the noise we found by LP-loop procedure of
\cite{CSLP} (looking for instanton configuration for the Linear
Programming decoding) has $l^2\approx 56.587$.

Another example is of the $(273, 191, 18)$ projective geometry code.
The connectivity degree for bits is $17$ thus instanton
configuration for one decoding step is $18$. With the number of
iterations increase the instanton length can not decrease, thus it
can not be less than $18$. On the other hand Hamming distance of the
code is exactly $18$. Therefore, the minimal instanton,
corresponding to median between two closest code words, has the
length of $18$.

\section{Future explorations}

Instead of standard Conclusions we briefly describe here further
directions for this research as we see it. The instanton-amoeba
technique is first of all a  practical tool we plan to perfect for
analysis of existed practical codes. Thus, it is important to extend
the list of long codes analyzed by this tool over various channel
models. Second, the instanton-amoeba, as a tool, will perform most
successfully if combined with other existed and emerging approaches
to the codes' analysis. An example of the complementary approach
which is capable of enhancing performance of the instanton-amoeba is
the LP-loop approach described in \cite{CSLP}. Third,
instanton-amoeba can help us to understand fundamental features of
the iterative decoding. In this context dynamical analysis of the
iterative decoding is another topic we are working on \cite{SC}.
Finally, the ultimate goal is to improve decoding and code design.
Improved and perfected instanton-amoeba is promised to be invaluable
and practical tool to achieve this big goal.

\section*{Acknowledgment}

The authors are grateful to V.~Chernyak, R.~Koetter, O.~Milenkovich,
T.~Richardson, and B.~Vasic for inspiring and fruitful discussions. This
work was supported by DOE under LDRD program at Los Alamos National
Laboratory.


\begin{thebibliography}{99}

\bibitem{04CCSV} V.~Chernyak, M.~Chertkov, M.G.~Stepanov, B.~Vasic,
{\it Error correction on a tree: an instanton approach}, {Phys. Rev.
Lett.} {\bf 93}, 198702 (2004).

\bibitem{05SCCV} M.G.~Stepanov, V.~Chernyak, M.~Chertkov, B.~Vasic,
{\it Diagnosis of weaknesses in modern error correction codes: a
physics approach}, Phys. Rev. Lett. {\bf 95}, 228701 (2005)
[extended version with supplemental materials --
arXiv.org:cond-mat/0506037].

\bibitem{05SC} M.G.~Stepanov, M.~Chertkov, {\it The error-floor of LDPC
codes in the Laplacian channel} --- 43rd Allerton Conference on
Communication, Control, and Computing (September 28--30, 2005,
Allerton House, Monticello, IL, USA) [arXiv.org:cs.IT/0507031].

\bibitem{63Gal} R.G.~Gallager, {\it Low density parity check
codes} (MIT Press, Cambridge, 1963).

\bibitem{99Mac} D.J.C.~MacKay,
{\it Good error-correcting codes based on very sparse matrices},
{IEEE Trans. Inf. Theory}~{\bf 45}, 399--431 (1999).

  \bibitem{04Ric} T.~Richardson, {\it Error floors of LDPC codes} ---
    41st Allerton Conference on Communication, Control, and Computing
    (October 1--3, 2003, Allerton House, Monticello, IL, USA).

\bibitem{05VK} P.~O.~Vontobel, R.~Koetter, {\it Graph-Cover Decoding
and Finite-Length Analysis of Message-Passing Iterative Decoding of
LDPC Codes}, arXiv.org:cs.IT/0512078 .

\bibitem{65NM} J.A.~Nelder, R.~Mead, {\it A simplex method for
    function minimization}, Computer Journal {\bf 7}, 308--313 (1965).

\bibitem{96W} N.~Wiberg, {\it Codes and decoding on general graphs},
    Ph.D. thesis, Link\"oping University, 1996.

\bibitem{SC} M.G.~Stepanov, M.~Chertkov, {\it Dynamics of iterative
    decoding}, in preparation.

\bibitem{01TSF} R.M.~Tanner, D.~Srkdhara, T.~Fuja,
%A Class of Group-Structured LDPC Codes,
{\it Proceedings of the 6th International Symposium on Communication
Theory and Applications, Ambleside, UK, July 15--20, 2001}, p.~365.

\bibitem{82M} G.A.~Margulis, {\it Explicit construction of graphs
wihtout short circles and low-density codes}, Combinatorica {\bf 2},
71--78 (1982).

\bibitem{02MP} D.J.C.~MacKay, M.J.~Postol, {\it Weaknesses of
Margulis and Ramanujan-Margulis Low-Density Parity-Check Codes},
Proceedings of MFCSIT2002, Galway.\\
http://www.inference.phy.cam.ac.uk/mackay/abstracts/margulis.html

\bibitem{CSLP} M.~Chertkov, M.G.~Stepanov, {\it Looping linear
    programming decoding of LDPC codes}, in preparation.

\end{thebibliography}
\end{document}